\documentclass[]{spie}  
\usepackage{makecell}
 
\usepackage{amsmath,amsfonts,amssymb}
\usepackage{graphicx}
\usepackage[colorlinks=true, allcolors=blue]{hyperref}
\usepackage{comment}

\title{GRAVITY+ Wavefront Sensors: High-Contrast, Laser Guide Star, Adaptive Optics systems for the VLTI}

\author[a]{Guillaume Bourdarot}
\author[a,b]{Frank Eisenhauer}
\author[a]{Şenol Yazıcı}
\author[a]{Helmut Feuchtgruber}
\author[d]{Jean-Baptiste Le Bouquin}
\author[a]{Michael Hartl}
\author[a]{Christian Rau}
\author[a]{Jonas Graf}
\author[a]{Nikhil More}
\author[a]{Ekkehard Wieprecht}
\author[a]{Frank Haussmann}
\author[a]{Felix Widmann}
\author[a]{Dieter Lutz}
\author[a,l]{Reinhard Genzel}
\author[c]{Frederic Gonté}
\author[c]{Sylvain Oberti}
\author[c]{Johann Kolb}
\author[c]{Julien Woillez}
\author[c]{Henri Bonnet}
\author[a]{Daniel Schuppe }
\author[a]{Amit Brara}
\author[a]{Johannes Hartwig}
\author[a]{Armin Goldbrunner}
\author[a]{Christoph Furchtsam}
\author[a]{Franz Soller}
\author[a]{Stefan Czempiel}
\author[a]{Johann Eibl}
\author[a]{David Huber}
\author[a]{Sinem Uysal}
\author[a]{Irmgard Treffler}
\author[a]{Hakan \"Ozdemir}
\author[a,b]{Vishaal Gopinath}
\author[c]{Pierre Bourget}
\author[h]{Anthony Berdeu}
\author[a]{Stefan Gillessen}
\author[a]{Thomas Ott}
\author[f]{Philippe Berio}
\author[f]{Olivier Boebion}
\author[f]{Florentin Millour}
\author[h]{Roderick Dembet}
\author[h]{Clémence Édouard}
\author[k]{Tiago Gomes}
\author[a]{Taro Shimizu}
\author[a]{Antonia Drescher}
\author[a]{Maximilian Fabricius}
\author[a]{Jinyi Shangguan}
\author[f]{Stéphane Lagarde}
\author[f]{Sylvie Robbe-Dubois}
\author[f]{Fatmé Allouche}
\author[d]{Hugo Nowacki}
\author[e]{Denis Defrère}
\author[k]{Paulo J. V. Garcia}
\author[j]{Sebastian Hoenig}
\author[g]{Laura Kreidberg}
\author[h]{Thibaut Paumard}
\author[i]{Christian Straubmeier}

\affil[a]{Max-Planck-Institut f\"ur extraterrestrische Physik, Gießenbachstraße 1, 85748 Garching bei M\"unchen, Germany}
\affil[b]{Department of Physics, TUM School of Natural Sciences, Technical University of Munich, 85748 Garching, Germany}
\affil[c]{European Southern Observatory, Karl-Schwarzschild-Str. 2, Garching bei M\"unchen, Germany}
\affil[d]{Universit\'e Grenoble Alpes, 621 Av. Centrale, 38400 Saint-Martin-d'H\`eres, France}
\affil[e]{Institute of Astronomy, KU Leuven, Celestijnenlaan 200D, 3001, Leuven, Belgium}
\affil[f]{Univ. Côte d’Azur, Observatoire de la Côte d’Azur, CNRS, Laboratoire Lagrange, France}
\affil[g]{Max-Planck-Institut für Astronomie, Königstuhl 17, 69117 Heidelberg, Germany}
\affil[h]{LESIA, Observatoire de Paris, Universit\'e PSL, Sorbonne Universit\'e, Universit\'e Paris Cit\'e, CNRS, 5 place Jules Janssen, 92195 Meudon, France}
\affil[i]{1st Institute of Physics, Universit\"at zu K\"oln, Z\"ulpicher Straße 77, 50937 K\"oln, Germany}
\affil[j]{University of Southampton, University Road, Southampton, SO17 1BJ, United Kingdom}
\affil[k]{Faculdade de Engenharia, Universidade do Porto, rua Dr. Roberto Frias, 4200-465 Porto, Portugal and
CENTRA - Centro de Astrof\'{\i}sica e Gravita\c c\~ao, IST, Universidade de Lisboa, 1049-001 Lisboa, Portugal}
\affil[l]{Departments of Physics and Astronomy, University of California, Berkeley, USA}

\authorinfo{Correspondence to: Guillaume Bourdarot (bourdarot at mpe.mpg.de)}

\begin{document} 
\maketitle

\begin{abstract}
We present the Wavefront Sensor units of the Gravity Plus Adaptive Optics (GPAO) system, which will equip all 8m class telescopes of the VLTI and is an instrumental part of the GRAVITY+ project. It includes two modules for each Wavefront Sensor unit: a Natural Guide Star sensor with high-order 40x40 Shack-Hartmann and a Laser Guide Star 30x30 sensor. The state-of-the-art AO correction will considerably improve the performance for interferometry, in particular high-contrast observations for NGS observations and all-sky coverage with LGS, which will be implemented for the first time on VLTI instruments. In the following, we give an overview of the Wavefront Sensor units system after completion of their integration and characterization.
\end{abstract}


\section{INTRODUCTION}
\label{sec:intro}
The GRAVITY+ project is a major upgrade of the GRAVITY instrument and of the VLTI facility, and will improve by orders of magnitude of sensitivity, contrast and sky coverage of GRAVITY [\citenum{Gravity2017}] and VLTI instruments as a whole [\citenum{Gplus2022}]. At the heart of the project is the implementation of Gravity Plus Adaptive Optics (GPAO). 
Early on in the development of interferometry, AO was identified as a crucial system: unlike small telescopes which are limited by the magnitude of fringe tracking, the limiting magnitude for large telescopes is set by the AO [\citenum{Eisenhauer2023}].
On large telescopes, for a correction set by the size of an atmosphere turbulent cell, a high-order correction is needed and the limiting magnitude is independent of the telescope diameter.
This results in a discrepancy between the fringe-tracking magnitude (in the infrared) which increases with telescope diameter, and the AO limiting magnitude (in the visible) which remains constant, translating in a constraint on the magnitude and the color limit on large telescopes.
For example, for an AO system with the same subaperture size as NAOMI, the typical color limit on the UTs is R - K = + 0mag, as estimated in [\citenum{Woillez2019}].
GPAO will lift the trade-off between high-order correction and limiting magnitude by implementing both a high-order Natural Guide Star (NGS) optimized for high-contrast on bright objects and a LGS mode on faint objects.
By doing so, GRAVITY+ achieves a long-term goal of implementing AO and Laser Guide Stars (LGS) for the VLTI, as envisioned from its initial design [\citenum{Beckers1990}].

\begin{figure} [ht]
   \begin{center}
   \begin{tabular}{c} 
   \includegraphics[height=8cm]{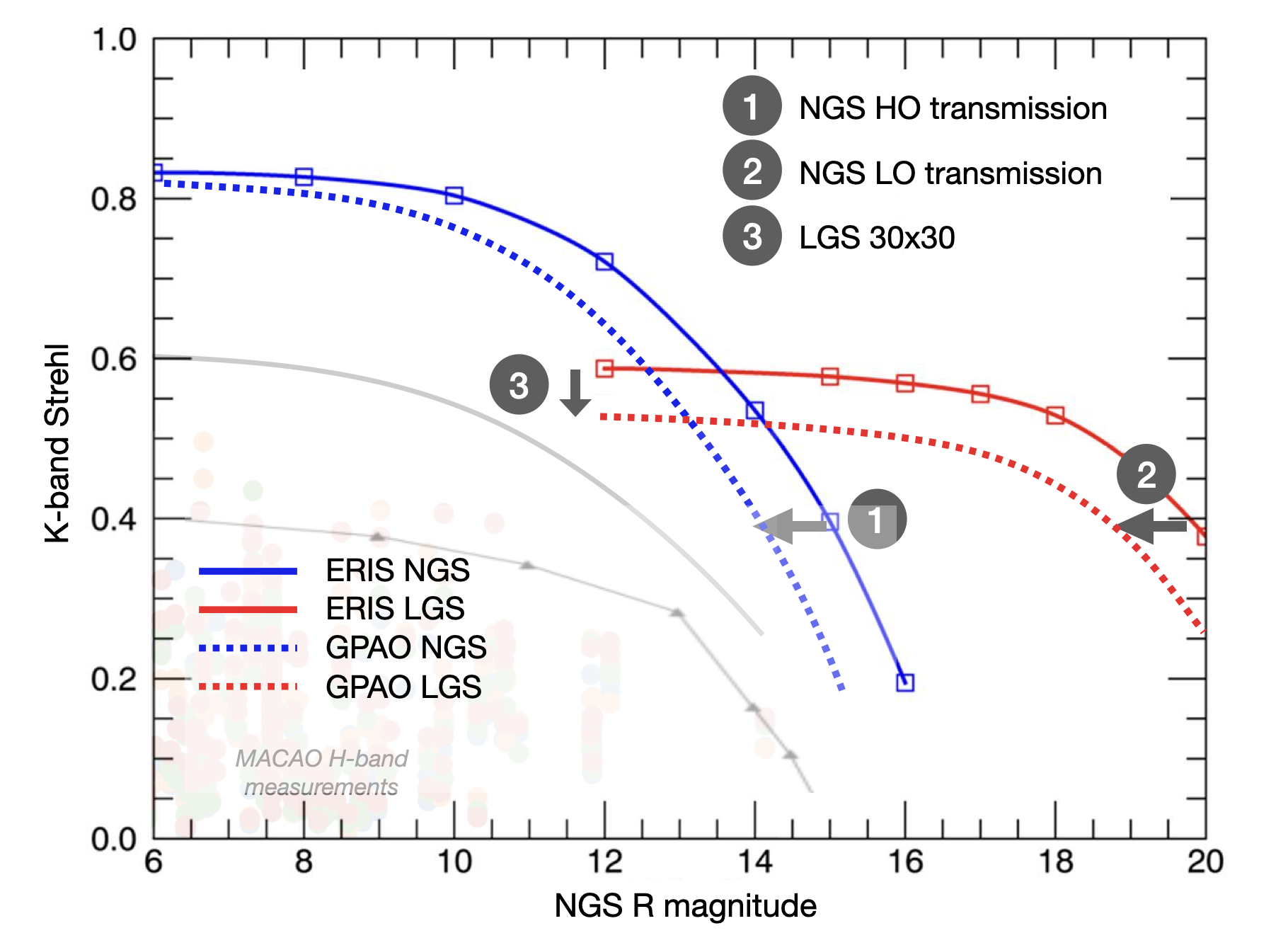}
   \end{tabular}
¨   \end{center}
   \caption[example] 
   { \label{fig:strehl} 
   Predicted Strehl ratio of the GRAVITY+ adaptive optics for NGS (blue dashed) and LGS modes (red dashed). The performances are adapted from ERIS, taking into account the additional throughpout losses of the UT Coudé train. MACAO on-sky performances are shown in the lower left corner (gray dots).}
   \end{figure} 

The GPAO NGS mode is designed for a limiting magnitude R=14 (Strehl=0.5), taking into account the losses of the Coudé train, and a Strehl 0.8 for bright objects. The LGS mode will offer a limiting magnitude R=19, set by the magnitude of the tip/tilt star.
The top-level requirement as well as the design of the GRAVITY+ WFS rely largely to the ones of ERIS [\citenum{Davies2023}], which was adapted to the UT Coudé focus.
The GPAO system will replace the MACAO curvature sensor [\citenum{Roddier1988,Arsenault2003}] in operation at VLTI for more than 20 years now. For faint objects, GPAO will increase by more than 5 magnitude the limiting magnitude of the AO star with the LGS compared to MACAO, and by a factor 2 - 2.5 the Strehl ratio for bright objects with NGS.
In addition, GPAO will remain compatible with the existing CIAO system, the 9x9 infrared sensor commissioned in 2016 with GRAVITY [\citenum{Scheithauer2016}], especially with a LGS+CIAO mode for objects for which no bright visible star is available. 


\section{SCIENCE DRIVERS}

\begin{figure} [h]
   \begin{center}
   \begin{tabular}{c} 
   \includegraphics[height=12cm]{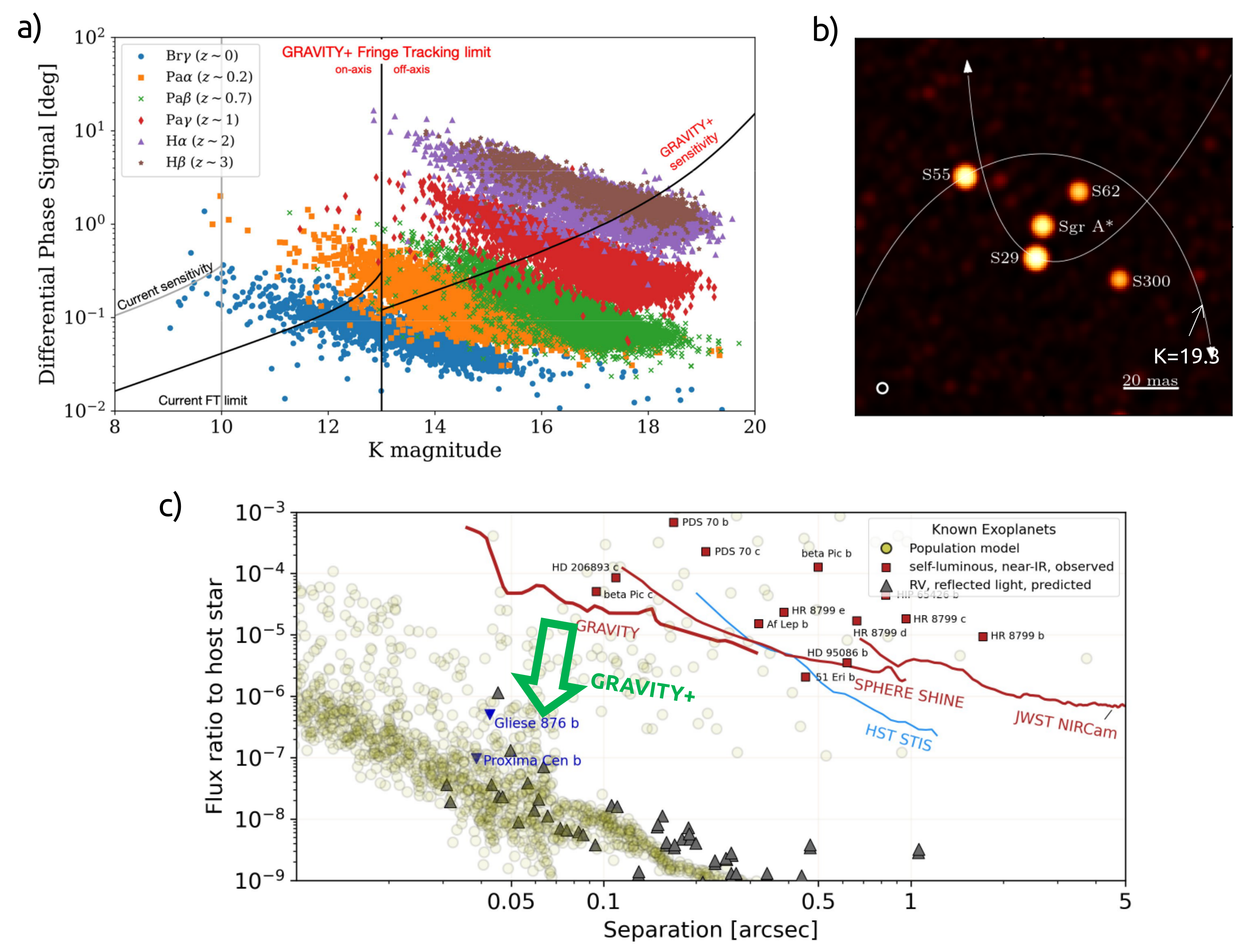}
   \end{tabular}
¨   \end{center}
   \caption[example] 
   { \label{fig:science} 
   \textbf{a)} High-redshift quasars: the LGS combined with the G-Wide upgrade will allow to increase the sample of currently 5 quasars to several dozens at redshift z=1-3. \textbf{b)} Galactic Center: the LGS+CIAO mode together with G-Faint upgrade will allow to go fainter than Kmag=21.0 for the imaging of the faint stars around the Galactic Center [\citenum{GC2022}], \textbf{c)} Exoplanets: the NGS 40x40 will allow the observations of exoplanets at high-contrast $10^{-6}$ and short separation $< 100 \text{mas}$ (adapted from [\citenum{Pourre2024}]).}
   \end{figure} 

\subsection{Supermassive Black Holes at Cosmic Noon}
\label{ref:SMBH}
The study of our local Universe indicates that there is a tight correlation between the evolution of galaxies and the mass of the Supermassive Black Holes (SMBH) which resides at their center. These objects are thought to co-evolve, despite the SMBH only contributes only to a small fraction of the total mass of the galaxy. 
In order to constrain this evolution, it is crucial to constrain the properties of SMBH across cosmic age.
Yet, our understanding is limited by our knowledge of SMBHs and in particular their mass, which relies on scaling relations using AGN reverbation mapping. Constraining the properties of SMBHs at high-redshift, while these objects are at the early stage of their formation, provides fundamental clues on how the black hole and its host galaxy evolve together.
GRAVITY+, using the off-axis tracking G-Wide [\citenum{GWide2022}], has demonstrated the direct measurement of SMBH dynamical mass at redshift z=2 through spatially resolved observations [\citenum{QSO2024}] of the Broad Line Region (BLR). Following the upgrade of the Beam Compressor Delay Lines (BCCDL, [\citenum{Fabricius2024}]), these observations have been recently expanded to 5 additional quasars. The GPAO upgrade, with the implementation of the LGS mode working together with G-Wide, will open up this sample to dozens of samples at redshift z=1 to 3, allowing for the first time to build a statistical sample of the SMBH mass at Cosmic Noon, during the peak of SMBH growth and star formation.

\subsection{Direct Characterization of Exoplanets}
The direct characterization of planets using GRAVITY the dual-field mode allows for the characterization of exoplanet atmosphere together with an astrometry accuracy of 50-100µas. 
The use of this technique has led to ground-breaking results, such as the first observation of an exoplanet with interferometry [\citenum{exoG2019}], the first direct characterization of a planet detected with radial velocity (RV) [\citenum{Nowak2020}], or the direct discovery of a GAIA and RV inferred planet [\citenum{Hinkley2023}].
Yet, these results were obtained with a moderate AO correction with a typical Strehl of 0.2 using MACAO.
Based on GPAO, the high-order correction with the NGS 40x40 will allow a dramatic improvement of the AO correction, translating both in better flux injection and reduced speckle noise. 
The improved AO correction will also allow the implementation of high-contrast wavefront control techniques, in particular to dig a dark hole at the location of the science fiber [\citenum{Por2020,Pourre2024}]. 
Together, these improvements will allow to push the contrast and inner-working angle of high-contrast observation to $10^{-6}$ at short separation $< 100 \text{mas}$ for the study of young planet in thermal emission.
These improvements open up the characterization of GAIA inferred planets, which will give access to the bulk of the population of young gas giant planets at 1 - 3 AU [\citenum{Fulton2021}].
Finally, the GPAO system will serve as a workhorse for all existing VLTI instruments and future high-contrast visitor instruments at VLTI [\citenum{Defrere2024}].

\subsection{Galactic Center}
The study of the Galactic Center has allowed precision tests of Einstein's theory of General Relativity, and delivered the strongest evidence that SgrA* is a Schwarschild-Kerr black hole. The next step is the measurement of the spin of the black hole [\citenum{Genzel2024}], which can be done by means of the Lense-Thiring precession of the orbit of surrounding stars. This requires the detection of a faint star on a close orbit around SgrA*.
The imaging of the Galactic Center is currently reaching a limiting magnitude of about $\sim$19 mag [\citenum{GC2022}]. 
This detection limit is set by the background flux of the stars around SgrA*, with a 1$\sigma$ average background noise of $21.0\pm 0.2$ mag in 3.25h integration time. 
In the case of the Galactic Center, this background originates from the bright surrounding stars, which can hardly be filtered out by the phase-referencing given the number of stars, which signal can hardly be modeled.
The CIAO+LGS will be a key gain in sensitivity, which will improve by a factor 2 the Strehl ratio from current S=25\% to S=50\% (Figure \ref{fig:strehl}). The improvement associated to the AO will be twofold: a direct gain of the injected flux by a factor 2, and the reduction of the noise by a factor 1.5 approximately by reducing the background flux (proportional to 1-S). Together with the reduction of the noise caused by the metrology laser [\citenum{Widmann2022}], and the upgrade of the Fringe-Tracker [\citenum{Nowak2024}], the total improvement of the performances should lead to a 3$\sigma$ detection limit in 3.25h of 22 mag in the Galactic Center.

\section{SYSTEM} 

\begin{figure} [ht]
   \begin{center}
   \begin{tabular}{c} 
   \includegraphics[height=10cm]{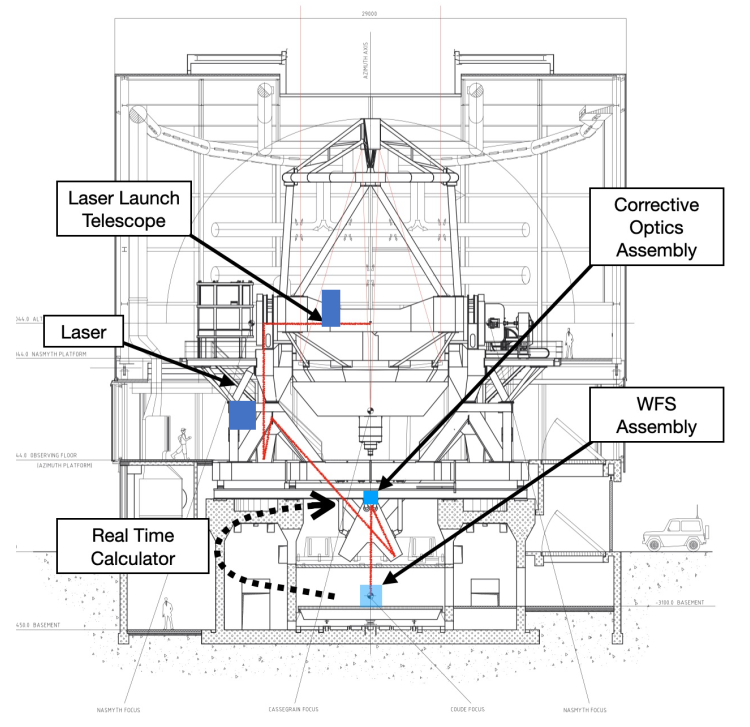}
   \end{tabular}
   \end{center}
   \caption[example] 
   { \label{fig:overview} 
Overview of the Gravity Plus Adaptive Optics subsystems.
}
   \end{figure}

The GPAO system is located in the Coudé Train of the UT and is composed of the following sub-systems (Figure \ref{fig:overview}):

\begin{itemize}
\item \textbf{Wavefront Sensors}: Shack-Hartmann 40x40 in NGS and 30x30 in LGS, located in the Coudé focus.
\item \textbf{Deformable Mirror}: an ALPAO 43x43 with 1.3k actuators located on M8, with an offload of tip/tilt to M2 guiding.
\item \textbf{Real Time Calculator} (RTC) based on SPARTA upgrade, located in the inner ring of the UTs
\item \textbf{Laser Guide Star}: is a high-power (20W) Raman fibre laser at sodium wavelength 589.158nm coupled to a telescope launcher located on the UT center piece. This subsystem is the same as the LGS on the Extremely Large Telescope.
\end{itemize}

\subsection{Wavefront Sensors}
The WFS are composed of a NGS and a LGS module based on Shack-Hartmann sensors. 
The NGS and LGS design was adapted from the ERIS AO [\citenum{Riccardi2016,Davies2023}] from a Cassegrain to a Coudé focus. 
The choice of Shack-Hartmann over a Pyramid sensor was driven by the need for linearity over a large range and the need of operational experience, which is lacking for the LGS with a Pyramid.
In the WFS unit, each module is mounted on a motorized XY-stage and can patrol the full Coudé field-of-view (FOV) of 2 arcmin. In operation:

\begin{itemize}
\item NGS mode is a High-Order (HO) 40x40 subapertures (with 6x6 pixels by subaperture) with a FOV of 2.5 arcsecond per subaperture.
\item LGS mode is a 30x30 subapertures (with 8x8 pixels by subaperture, FOV of 5 arcsec) working with the Low-Order (LO) 4x4 on the NGS module for the tip/tilt star. The HO-LO switch allows to use on the same NGS camera.
\end{itemize}

The observing modes of GPAO will be the following (see [\citenum{Berdeu2024}]):

\begin{table}[ht]
\label{tab:fonts}
\caption{Observing modes of the Gravity Plus Adaptive Optics System.}
\begin{center}     
\begin{tabular}{|c|c|c|c|c|}
\hline 
Mode & NGS VIS & LGS VIS & NGS IR & LGS IR \\ 
\hline 
SH Sensor & NGS-HO: 40x40 & \makecell{LGS: 30x30\\ NGS-LO: 4x4 (tip/tilt)} & CIAO: 9x9 & \makecell{LGS: 30x30 \\ + CIAO: 9x9 (tip/tilt)} \\ 
\hline 
Pixel per subap & 6 & LGS: 8 / LO: 12 & CIAO: 8 & LGS: 8 / CIAO: 8 \\ \hline
Field-of-View & 2.5 as & 5 as & 2.0 as & 5 as \\ \hline
\end{tabular}
\end{center} 
\end{table} 

The WFS are located after M8 (Deformable Mirror), M9 dichroic beam splitter which separates the visible light (AO) from the infrared light sent to VLTI, and the telecentric lens for the patrol field. GPAO-WFS modules are placed below the Star Separator (STS), at the same position as the MACAO modules. The opto-mechanical design and the positioning in the mechanical frame of the UT Coudé focus can be found in [\citenum{More2024}] for a detailed description.

\begin{table}[ht]
\caption{Design of the Natural Guide Star and Laser Guide Star sensors.}
\label{tab:fonts}
\begin{center}       
\begin{tabular}{|c|c|c|} 
\hline
\rule[-1ex]{0pt}{3.5ex}   & NGS & LGS \\
\hline
\rule[-1ex]{0pt}{3.5ex} Wavelength Range  & 600nm - 1000nm & 569nm - 609nm \\
\hline
\rule[-1ex]{0pt}{3.5ex} Shack-Hartmann  & HO: 40x40 / LO: 4x4 & 30 x 30 \\
\hline
\rule[-1ex]{0pt}{3.5ex} Field of View  & Ø 2.5 as & Ø 5.0 as \\
\hline
\rule[-1ex]{0pt}{3.5ex} Patrol Field  & Ø 2.0 arcmin & Ø 2.0 arcmin \\
\hline
\rule[-1ex]{0pt}{3.5ex} Focus Range  & - & 80 km to $\infty$ \\
\hline
\rule[-1ex]{0pt}{3.5ex} Focal Ratio  &  \multicolumn{2}{c|}{f/47.0} \\
\hline
\rule[-1ex]{0pt}{3.5ex} Camera  & \multicolumn{2}{c|}{OCAM2 up to 2kHz, 240x240 pixels } \\
\hline
\end{tabular}
\end{center} 
\end{table}

\subsection{Natural Guide Star}

\subsubsection{NGS Optical Design}
\begin{figure} [b]
   \begin{center}
   \begin{tabular}{c} 
   \includegraphics[height=7cm]{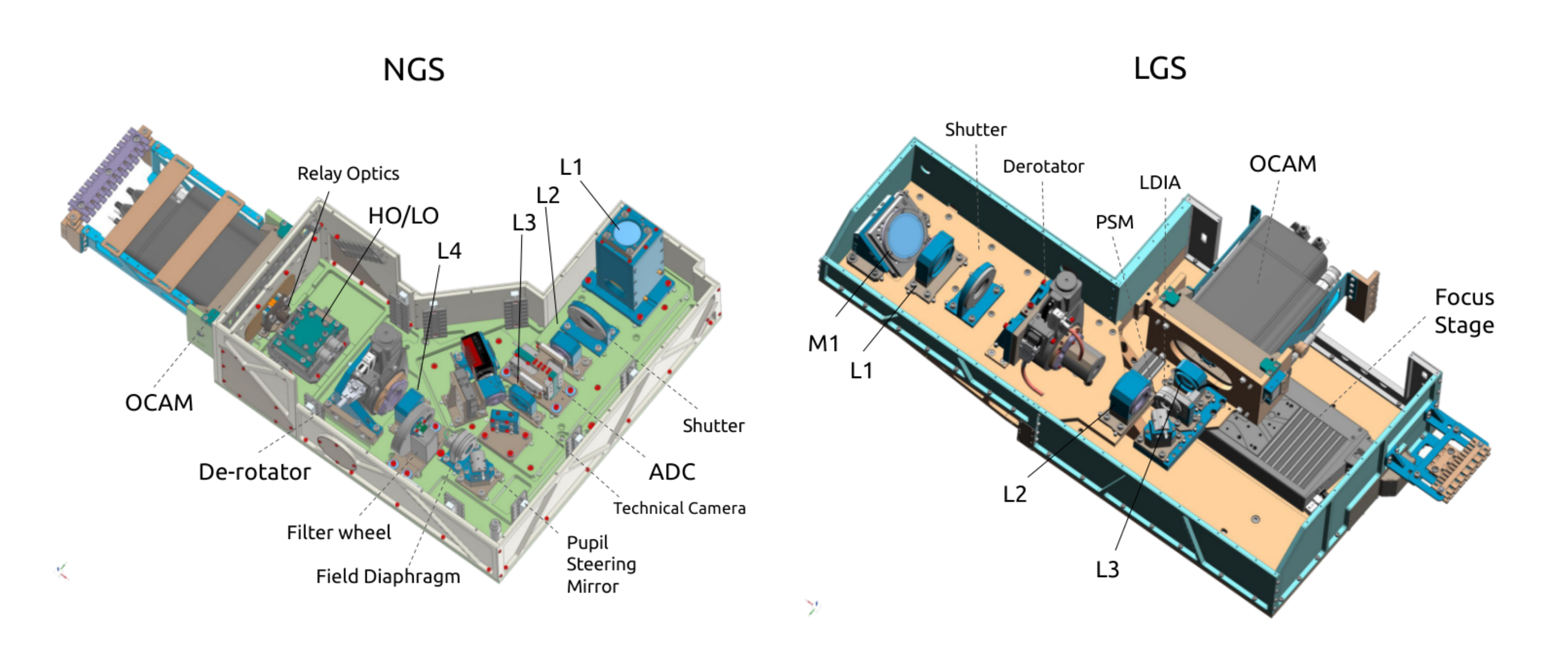}
   \end{tabular}
   \end{center}
   \caption[example] 
   { \label{fig:WFS} 
\textbf{Left}: Natural Guide Star wavefront-sensor module \textbf{Right:} Laser Guide Star sensor.
}
   \end{figure}

\begin{figure} [b]
   \begin{center}
   \begin{tabular}{c} 
   \includegraphics[height=7cm]{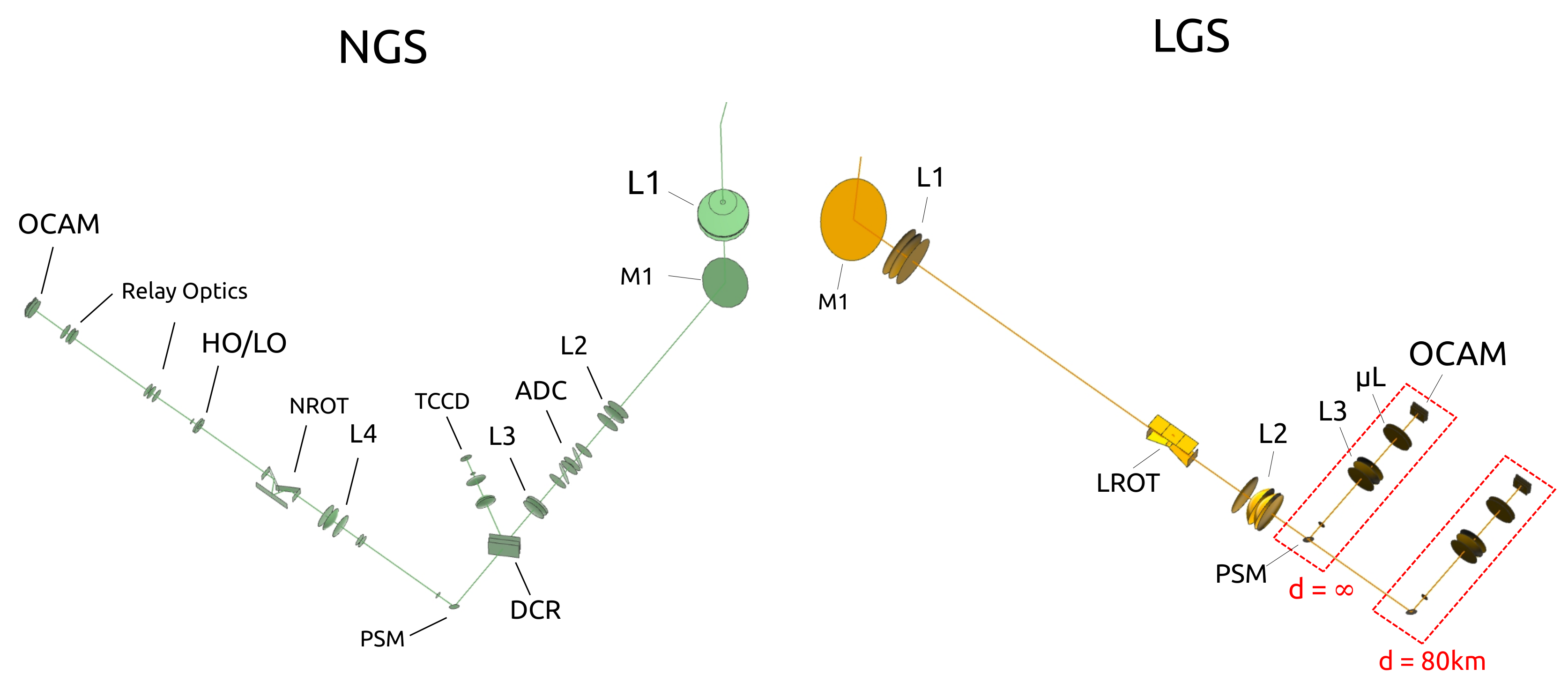}
   \end{tabular}
   \end{center}
   \caption[example] 
   { \label{fig:WFS_layout} 
Optical Layout of the NGS and the LGS modules.
}
   \end{figure}

The L1 lens of the NGS is located close to the Coudé focus and images the exit pupil of the telecentric lens to the ADC. The lens L2 (located just before the ADC) and L3 (right after the ADC) creates a f/20 beam on a focal plane at the Pupil Steering Mirror (PSM) and the Technical CCD (TCCD). After the PSM, the diaphragm provides an adjustable field stop. After the filter wheel, the lens L4 re-images the pupil, which can be de-rotated by the K-mirror, on the lenslet of the Hi-Lo translation stage. The HO (40x40) and LO (4x4) are mounted on the Hi-Lo stage which allows to select the mode of observations of the NGS. The pupil is finally re-imaged by a relay optics onto the e2v CCD220 chip. The optical layout of the NGS can be found on Figure \ref{fig:WFS} and Figure \ref{fig:WFS_layout}.
   
\subsubsection{NGS Subsystems}

\paragraph{NXYT Translation stage}
The full NGS module is mounted on a XY-translation stage, which allows to cover the 2.0 arcmin patrol-field. The micrometer precision of the stage allows for fine adjustement in the field, as L1 is located close to the Coudé focus (pixel scale 0.549 as/mm at the Coudé focus). In tracking mode, the NXYT allows to follow a given trajectory within the patrol field.

\paragraph{Atmospheric Dispersion Compensator (ADC)}
The ADC consists of a pair of counter-rotating Amici prisms, based on combination of S-FPM2 and Schott FK5 glasses. The latter is replacing the S-TIM8 in the ERIS design. The ADC is located in a pupil plane of the NGS, between L2 and L3 lenses. It is optimized for a spectral range of 600-1000nm, and is designed for a maximum Zenith angle of 70$^{\circ}$ assuming the environment data in terms of height, latitude, humidity of Paranal observatory.

\paragraph{Technical Camera (TCCD)}
The NGS module includes a technical acquisition camera, based on a Prosilica GT 2050 camera. The camera is placed after a dichroic beam-splitter (DCR) located between L3 and the PSM, which reflects the blue part of the spectrum 400-570 nm ($R>95\%$) to the TCCD, while transmiting the beam necessary for the OCAM (589-1000 nm, $T>99\%$). The TCCD is equipped with a lens (NPV) which can be slid in and out of the beam to image the pupil or the field respectively.

\paragraph{Pupil Steering Mirror (PSM) and Field Diaphragm (NDIA)}
The PSM is located in a focus plane with f/20.0, and allows to stabilize the pupil on the micro-lens. This device is based on a piezo-actuator PI S-334 and allows for good repeatability and travel range. The PSM is located just upstream from a motorized iris which can be used as an adjustable field diaphram (NDIA). The diaphram can be adjusted from 0 to a maximum 8 arcsec approximately, which allows to control the FOV of the subaperture and adapt it to the seeing conditions. 

\paragraph{K-mirror derotator (NROT)}
The baseline is to operate the WFS without a K-mirror using a software derotation based on a synthetic model of the system, as implemented in []. This model was updated to take into account magnification and anamorphoses, which are known to be important in the CIAO system. Nonetheless, as part of the risk mitagation strategy during the design of GPAO, a derotator was implemented. The K-mirror is made of a FK5 glass, which provide low chromatic dispersion. The dimensions of the K-mirror are such that they allow removing this element withouth changing the overall optical path length. Alternatively, this element can also be used for calibration purpose to generate field rotation.

\paragraph{Calibration Unit (NCAL)}
The WFS also embedds a calibration unit, which consists in a halogen lamp. The lamp is connected to a multimode fiber, which can be placed in or out of the beam, at the position of the UT Coudé focus just above L1. The calibration unit can therefore be used to generate a point-like source in the field. This calibration mode is used for the daily characterization of the WFS and for monitoring its basic parameters (Detector gain, internal Wavefront Error, Pixel Scale, Shack-Hartmann field-of-view), see \ref{sec:parameters}.

\subsection{Laser Guide Star}
\subsubsection{LGS Optical Design}
The LGS beam is reflected by the dichroic located below the telecentric lens towards the LGS arm. The LGS focus is thus located between the dichroic and the Periscope Folding Mirror (PFM). The beam is then converted from a f/47 to a f/12 beam by the lenses L1 and L2, in order to reduce the focusing range from 1.7m (longitudinal shift at 80km for a f/47 beam) to 120mm. Between L1 and L2, the K-mirror is located close to a pupil plane and allows for field derotation. The focus is ajusted by a translation stage (LFOC), on which the downstream WFS optics are mounted (from L3 to the detector). Like in the NGS, the field is re-imaged on the PSM with an adjustable field stop located in its vicinity.  Finally, the pupil is re-imaged by L3 on the 30x30 lenslet array, which is directly mounted in front of the CCD chip (unlike the NGS, the lenslet in the LGS does not have to move). The optical layout of the LGS is shown on Figure \ref{fig:WFS} and \ref{fig:WFS_layout}.

\subsubsection{LGS Subsystems}

\paragraph{XY Translation stage (LXYT) and Focus Stage (LFOC)}
The LGS module is mounted on a XY-Translation stage, which allows to cover the patrol field in the same way as the NGS. In addition, the LGS payload comprises a focus stage, which allows to adjust for a laser guide star ranging from 80km to infinity. The LFOC can be placed in tracking mode, which automatically update the position based on the value of the focus from the truth sensor and the altitude of the telescope.

\paragraph{LGS actuators}
The control strategy and the actuators in the LGS module are similar to the NGS, which comprises a pupil steering mirror (LPSM), a field-diaphragm (LDIA), a K-mirror (LROT) and a calibration unit (LCAL) at the position of the LGS Coudé focus. The actuators and these components are the same as the NGS, with the optics adapted to the spectral range of the LGS (569-609nm).

\begin{table}[ht]
\caption{Specifications of the Wavefront Sensor units.}
\begin{center}
\begin{tabular}{|c|c|c|}
\hline 
Parameter & Method & Criteria \\ 
\hline 
OCAM Noise & \makecell{Noise level in dark conditions \\ at gain 1 and 800 (with fps 100 Hz and 1 kHz)} & \makecell{250 e-/s at gain 1 \\ 0.4e-/s at gain 800}  \\ 
\hline 
WFE & Wavefront Error measurement on the Shack-Hartmann & $<150 nm$ RMS \\ 
\hline 
Pixel Scale & \makecell{Equal flux between two neighboring pixels,\\ scanning the XYT to move the SH spots} & \makecell{HO: 0.777 mm/pix \\ LO: 0.394mm/pix\\ LGS:1.480mm/pix} \\ 
\hline 
Field of View & \makecell{Size and vignetting of the FOV by scanning the XYT} & No Vignetting over FOV \\ 
\hline 
K-mirror wobble & \makecell{Wobble of the FOV for a mechanical 360$^{\circ}$ K-mirror rotation} & $<2$ subap\\ 
\hline 
ADC wobble & \makecell{Wobble of the FOV for a mechanical 360$^{\circ}$ ADC rotation} & $<0.2$ as\\ 
\hline 
\end{tabular} 
\end{center}
\end{table}

\section{PERFORMANCE}

\subsection{Functional tests \& Performance}
\label{sec:parameters}
The functional tests of the WFS were validated in Europe, both at the unit level (WFS, Garching) and at System level (Nice, see [\citenum{Millour2024}]). 
This sequence was automated through VLT templates (see \ref{sec:templates}) delivered with the instrument, and can be run as a daily health-check. All four WFS units were characterized, and the results were validated during the Preliminary Acceptance Europe (PAE) at ESO in May 2024. The Wavefront Sensor units have been shipped to Chile in May and June 2024, are being commissioned at Paranal in Summer 2024.

\begin{figure}[ht]
   \begin{center}
   \begin{tabular}{c} 
   \includegraphics[height=7.5cm]{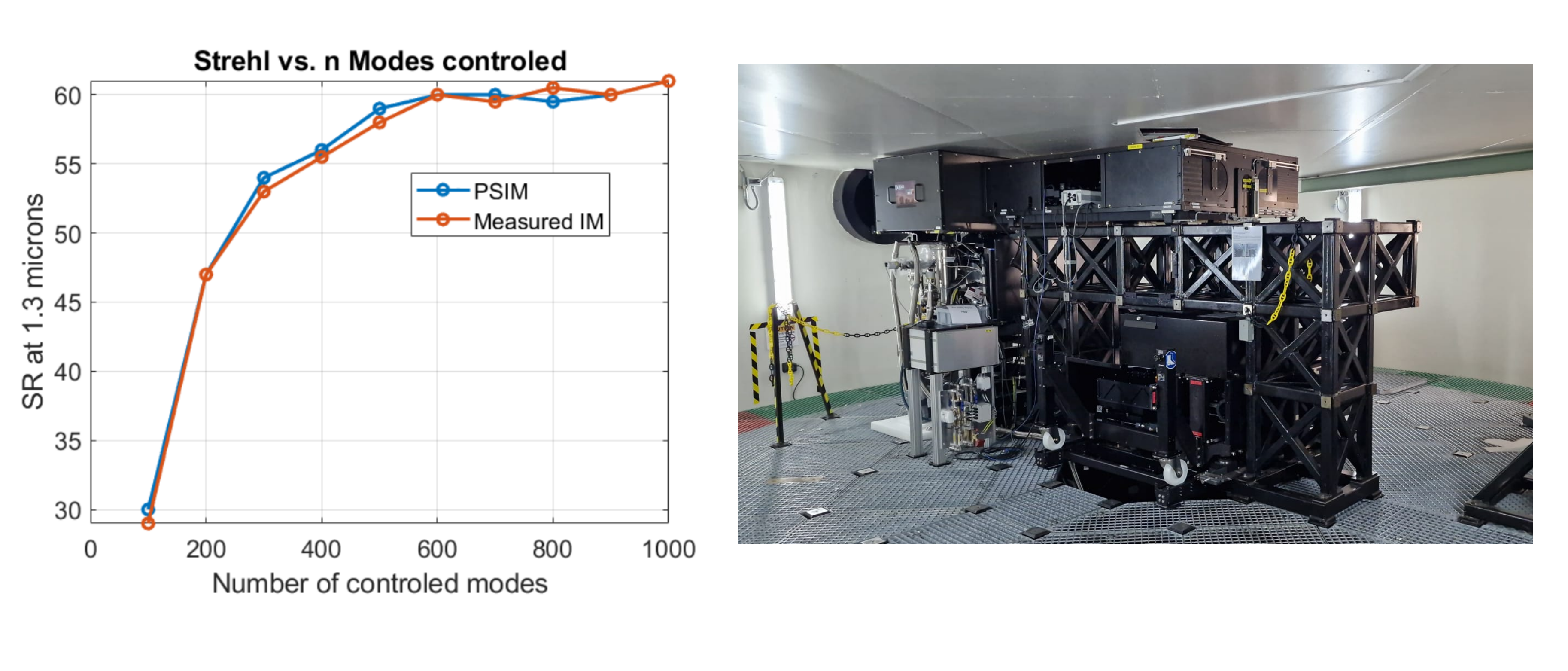}
   \end{tabular}
   \end{center}
   \caption[example] 
   { \label{fig:perfs} 
\textbf{Left}: Strehl ratio as a function of the number of controlled modes. The AO system shows no decrease of performances up to a number of controled mode of 1000, and shows a good agreement between the measured Interaction Matrix (IM) and the pseudo-Interaction Matrix (PSIM). \textbf{Right:} View of one GRAVITY+ WFS unit installed at the Coudé focus of UT1 in Paranal.}
\end{figure}

The performance of the WFS was validated on the NGS closing the AO loop while controlling up to 1000 modes, without losses of performance (Figure \ref{fig:perfs}). The integral gain $K_i$ of the NGS reaches a plateau of performance between 0.2 and 0.7. Both aspect leaves margin for the baseline mode, with 500 modes corrected and an integral gain $K_i=0.5$. For the LGS, similar performances were obtained, with an optimal gain between $K_i=$ 0.2 and 0.5 and no performance losses up to 500 modes. The baseline mode for the LGS is currently 400 modes corrected and an integral gain of 0.3. These values are being confirmed on-sky with the commissioning of the GPAO modules at the telescopes as these lines are being written.

\begin{figure}[t]
   \begin{center}
   \begin{tabular}{c} 
   \includegraphics[height=8.0cm]{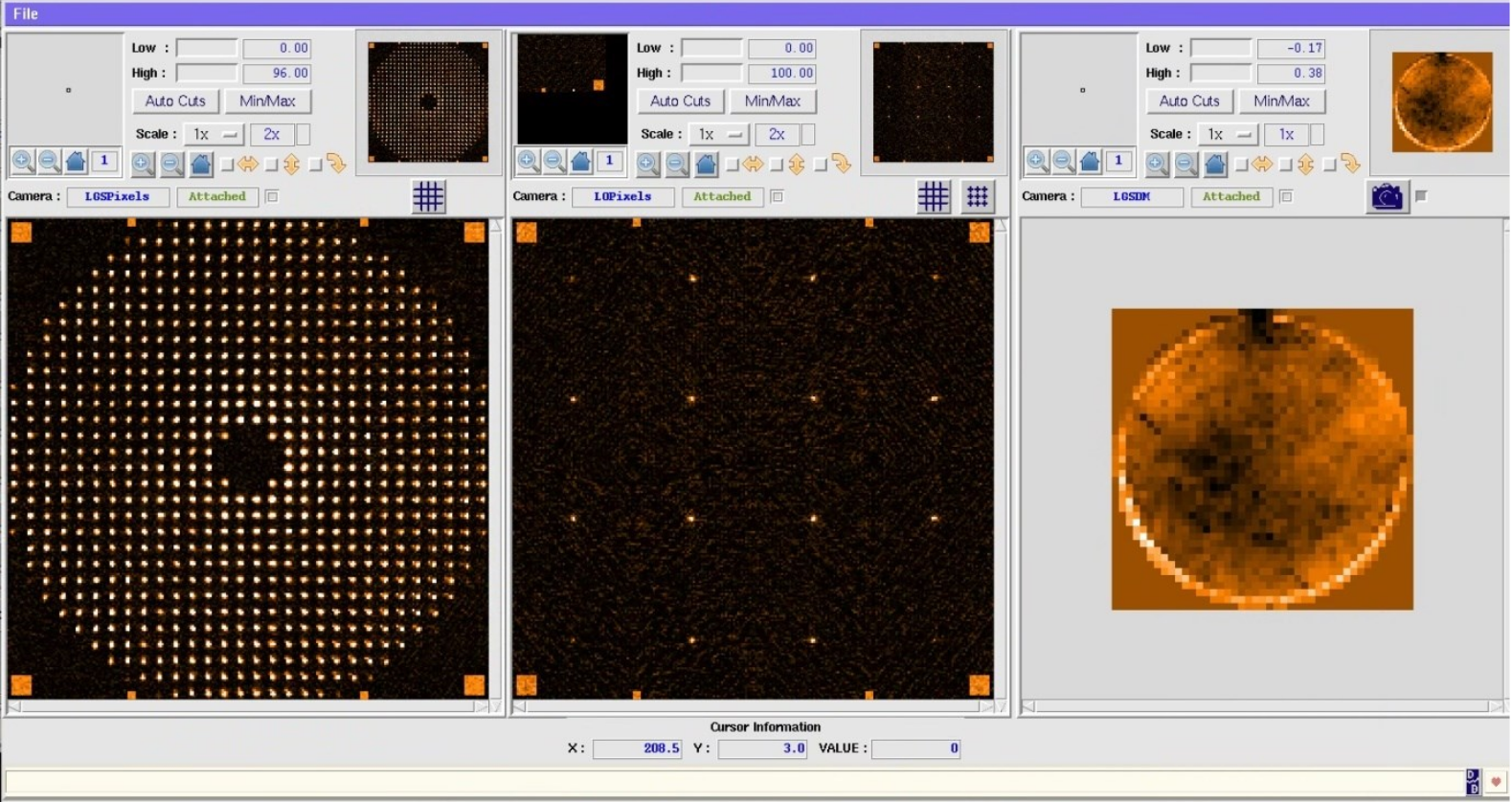}
   \end{tabular}
   \end{center}
   \caption[example] 
   { \label{fig:LGS_overview} 
Real-Time display in LGS mode with the 30x30 LGS wavefront sensor (left), the NGS wavefront sensor 4x4 Tip-Tilt sensor (center) and the commands applied to the DM (right). In NGS-VIS mode with 40x40 High-Order (not shown), only the NGS sensor is used and the 4x4 is automatically replaced with a 40x40 lenslet.}
\end{figure} 

\subsection{OCAM2 camera}
The NGS and LGS modules both rely on the OCAM2 camera from First Light Imaging. The cameras are based on an e2v EMCCD which can be operated full-frame up to 2 kHz with a multiplication gain up to 1000. The read-out noise of the cameras was characterized in the lab and is equal to 0.3e/s on average, for a multiplication gain of 800 and running at 1kHz. The multiplication gain of the camera is known to drift per octant over a period of a few month. Therefore the multiplication gain is monitored over time using the internal calibration unit and corrected through the voltages applied to each octant, in order to keep the homogeneity of the gain on the chip.

\subsection{Python Templates in VLT2022}
\label{sec:templates}
The WFS modules were integrated and delivered with templates developed in Python within the VLT2022 framework. These templates are used to perform the characterization tests and the daily health check of the modules. These templates can all run independently of SPARTA (e.g. commissioning purposes, maintenance, etc.), and are complimentary to the templates at system-level presented in [\citenum{Millour2024}].

\subsection{Integration}
At the time of writing this proceeding, the WFS modules have been shipped and are being integrated in Paranal.
The integration and alignment of the WFS modules in Garching were performed using a so-called telescope simulator, which allows to generate a telecentric beam, with either point-source in a field plane with an adjustable pupil stop, or a point in a pupil plane with an adjustable field stop. This source is aligned on a reference target attached to the tower structure, and which reproduces the tilt and centering position of the UT beam. The telescope simulator also includes a sighting telescope coaligned with the reference target, and serves as a reference axis for alignment. In the UTs, a similar set-up is reproduced with a sighting telescope in the STS, aligned on the azimuth axis of the telescope and using the Nasmyth beacon of the UTs.

\section{CONCLUSION}

The four GPAO units have been assembled and characterized in Europe, and shipped to Paranal in June 2024. The implementation of state-of-the-art AO at the VLTI will allow for the first time high-contrast observations with the NGS mode, and will be followed by the commissioning in 2025 of the LGS guide stars with global sky coverage, and with a significant increase of operability during average seeing conditions compared to MACAO. 
It completes a long-term vision initiated during the design of the VLTI, and breaks a fundamental sensitivity limitation of interferometry, entering the era of high-contrast observations and all-sky coverage with interferometry.

\acknowledgments 
 
GRAVITY+ is a consortium composed of German (MPE, MPIA, University of Cologne), French (CNRS-INSU:
LESIA, Paris, IPAG, Grenoble, Lagrange, Nice, CRAL, Lyon), British (University of Southampton), Belgian
(KU Leuven), and Portugese (CAUP) institutes, built in close collaboration with ESO. DD has received funding from the European Research Council (ERC) under the European Union's Horizon 2020 research and innovation program (grant agreement CoG - 866070). TP has received funding from the European Union’s Horizon 2020 research and innovation program under grant agreement No 101004719.

\bibliography{report} 
\bibliographystyle{spiebib} 

\end{document}